\documentclass[twocolumn,groupedaddress,amsmath,amssymb,pra]{revtex4-2}
\usepackage{graphicx}%
\usepackage[pdftex,colorlinks=true,linkcolor=blue,citecolor=blue,urlcolor=blue]{hyperref}
\newcommand{\ket}[1]{\left| #1 \right>} 

\begin{document}
\title{Fine-grained topological structures hidden in the Fermi sea}
\author{Wei Jia}
\email{jiaw@lzu.edu.cn}
\affiliation{Key Laboratory of Quantum Theory and Applications of MoE, Lanzhou Center for Theoretical Physics, and Key Laboratory of Theoretical Physics of Gansu Province, Lanzhou University, Lanzhou 730000, China}

\begin{abstract}
The geometry of the Fermi sea hosts a unique form of quantum topology, playing an equally significant role as the geometry of wave function. It governs the conductance quantization of a metal and is characterized by the Euler characteristic $\chi_F$, offering a new perspective in the study of topological quantum matter. Here, we reveal a class of fine-grained topology in the Fermi sea, whose characterization goes beyond $\chi_F$. Such topological structure in the symmetry-protected Morse-function regime leads to a nontrivial result that two Fermi seas with identical $\chi_F$ but different fine-grained topologies cannot be connected without a Lifshitz transition. To encode this structure, we introduce a structural resolution factor, which exactly captures the deeper topological information of the Fermi sea. When further considering the attractive Hubbard interaction of electrons on Fermi surfaces, we demonstrate that the resulting chiral topological superconducting phases can inherit the fine-grained Fermi sea topology of their parent metallic bands, with differences in these fine-grained structures giving rise to anomalous Majorana boundary states at the interface between two metal/superconductor heterojunctions. This work opens an avenue for exploring the topological richness of the Fermi sea.
\end{abstract}
\maketitle

\section{Introduction}~Quantum topology, as a cornerstone of modern condensed matter physics, reshapes our understanding for electronic phases and material properties~\cite{klitzing1980new,PhysRevLett.48.1559,hasan2010colloquium,qi2011topological,RevModPhys.89.041004}. A very prominent example is topological phases~\cite{laughlin1981quantized,thouless1982quantized,chiu2016classification}, whose quantum topology arises from the global feature of wave function across the Brillouin zone (BZ), determined by defining topological invariant of the ground state~\cite{kitaev2009periodic,PhysRevB.83.075103,PhysRevLett.106.106802,morimoto2013topological,PhysRevB.90.165114,ren2016topological,PhysRevLett.132.083801,PhysRevLett.134.163801}. In a metal, however, there is another quantum topology dictated by the geometry of the Fermi sea, which impacts the quantized response of the system~\cite{landauer1957spatial,fisher1981relation,buttiker1986four,stone1988measured}. 
The associated conductance quantization has been observed in quantum point contacts~\cite{van1988quantized}, semiconductor nanowires~\cite{honda1995quantized,van2013quantized}, and carbon nanotubes~\cite{frank1998carbon}. Recently, a breakthrough has elucidated that such Fermi sea topology is characterized by the Euler characteristic $\chi_F$~\cite{kane2022quantized}. With this framework, various probing schemes of the Fermi sea topology have been proposed, including multipartite entanglement~\cite{tam2022topological}, Andreev state transport~\cite{tam2023probing,tam2023topological}, and density correlations of Fermi gas~\cite{tam2024topological,daix2025probing,tam2026singular}. The quantized response may also possibly be measured in an ultracold atomic gas~\cite{yang2022quantized,zhang2023quantized,kfwv-7wh4}.

Despite its long-standing importance, the Fermi sea  topology remains fundamentally elusive. The main challenge lies in a fact that certain Lifshitz transitions (LTs) go beyond the characterization of $\chi_F$~\cite{volovik2018exotic}. This difficulty also applies to exotic Fermi seas with M\"obius Fermi surfaces~\cite{PhysRevLett.130.066001} and fractional Fermi seas~\cite{zeng2026realization,bastianello2026exotic}. This means that the current understanding of the Fermi seas topology remains incomplete. Hence, it is urgently  necessary to uncover the topological structures beyond the characterization of $\chi_F$, which is of crucial importance for a comprehensive understanding of the Fermi sea topology. Furthermore, the electrons on Fermi surfaces (FSs) of metallic bands can produce Cooper pairs by considering an attractive Hubbard interaction, thereby inducing the emergence of topological superconducting (SC) phases~\cite{PhysRevLett.100.096407,liu2014realization,poon2018semimetal,jia2019topological}. The properties of Majorana boundary states are closely related to the Fermi sea topology of their parent metallic bands~\cite{yang2023euler,PhysRevB.111.155115}. Therefore, it is extremely interesting to reveal new physical effects in topological superconductors caused by the Fermi sea topology characterized far from $\chi_F$.

\begin{figure}[t]
\centering
\includegraphics[width=1.0\columnwidth]{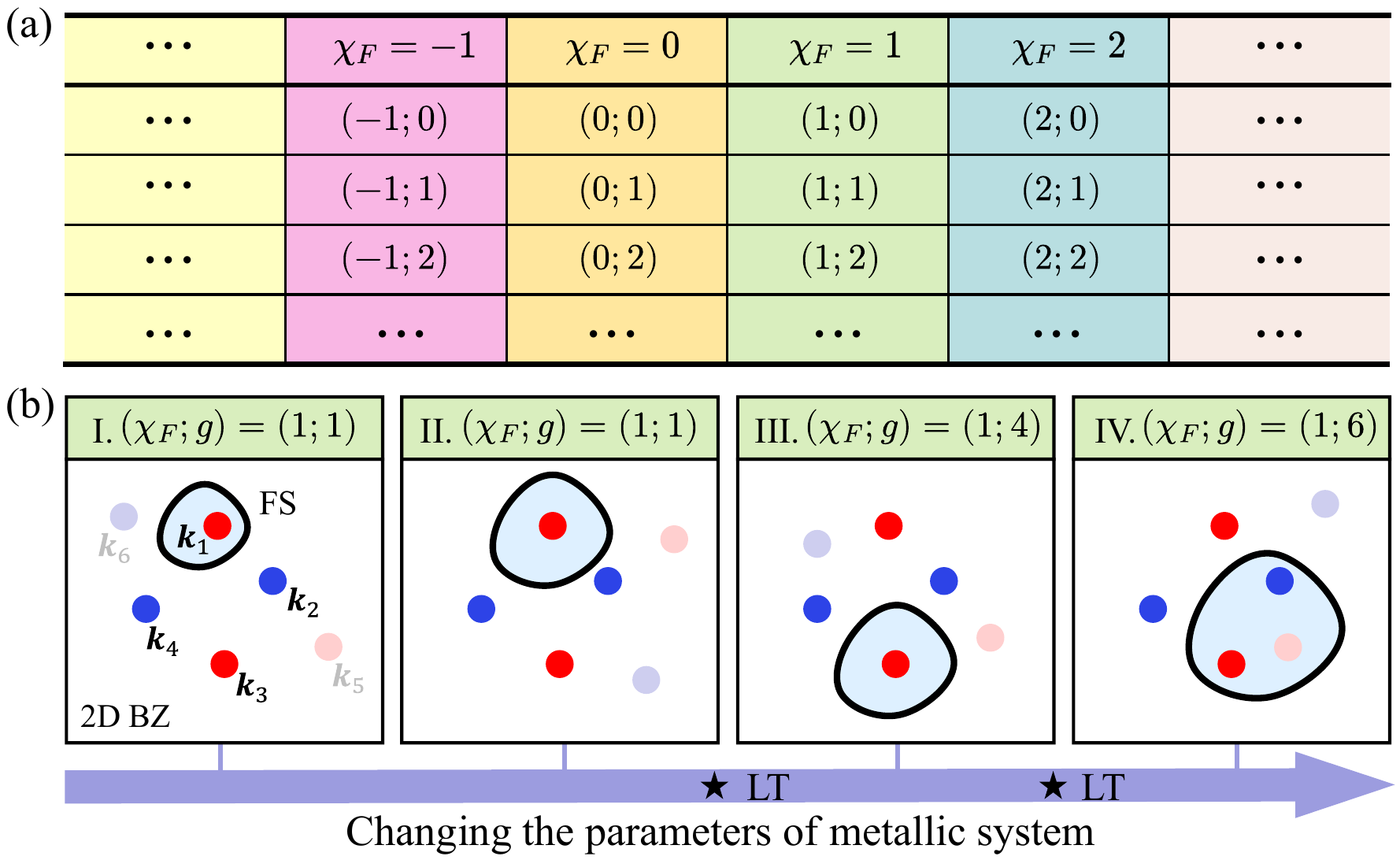}
\caption{(a) Topological indices $(\chi_F;g)$ provide a comprehensive description of Fermi sea topology, where the structural resolution factor $g$ determines its fine-grained topological structures. (b) A schematic to calculate $(\chi_F;g)$ for four different 2D Fermi seas. Here $\mathbf{k}_{1,2,\cdots,6}$ are critical points hosting $\eta_{1,3,5}=1$ and $\eta_{2,4,6}=-1$, where $\mathbf{k}_{1,2,3,4}$ are the FCPs and $\mathbf{k}_{5,6}$ are the ACPs. The fine-grained topological structures of I (II), III, and IV are markedly different.}
\label{fig:1}
\end{figure}

In this paper, we discover a class of fine-grained topology in the Fermi sea, showing that two-dimensional ($2$D) Fermi seas with identical $\chi_F$ in the symmetry-protected Morse-function regime cannot be connected without passing through a LT if their fine-grained topological structures are different. We further introduce a structural resolution factor to establish a completely theoretical framework to describe the Fermi sea topology, revealing a layer of topological information beyond the reach of $\chi_F$. Moreover, we show that chiral topological SC phases emerging from attractive Hubbard interactions of electrons on FSs can definitely inherit the fine-grained topology of their parent metallic bands. The differences of these structures give rise to anomalous Majorana boundary states at the interface between two metal/superconductor heterojunctions with identically topological systems, which challenges the traditional understanding that the Majorana interface states only appear at the interface between topologically distinct systems. Our work deepens the understanding of Fermi sea topology and provides an insight for exploring the topological richness in the Fermi sea.

\section{Fine-grained Fermi sea topology}~Our starting point is a $2$D metallic system featuring the electronic band $E_\mathbf{k}$ in momentum space. The Fermi sea in such a system gives rise to a quantum topology characterized by $\chi_F$, which governs its quantization of nonlinear conductance~\cite{kane2022quantized}. The Morse theory~\cite{milnor1963morse,nash1988topology} provides a direct way to calculate $\chi_F$ through  
\begin{equation}
\chi_F=\sum_{m}\eta_m.
\end{equation}
Here $m$ labels the critical points $\mathbf{k}_m$ within the Fermi sea, where $\mathbf{v}_\mathbf{k}=\nabla_\mathbf{k}E_{\mathbf{k}}=0$ and $E_\mathbf{k}<E_F$. For convenience, the Fermi level $E_F$ is taken as zero energy. The signature of each critical point is given by $\eta_m=\text{sgn}[\text{det}(\mathbb{H})]$, where $\mathbb{H}$ is the Hessian matrix of $E_\mathbf{k}$. Note that $\mathbf{k}_m$ are considered to be nondegenerate, i.e., $\text{det}(\mathbb{H})\neq 0$. When a critical point passes through $E_F$, it allows $\chi_F$ change at a LT. Under this theoretical framework, two Fermi seas are topologically equivalent provided one can be smoothly deformed into the other without undergoing a LT~\cite{lifshitz1960anomalies}, since they share the same $\chi_F$. Nevertheless, if the deformation involves LTs that change the number and position of $\mathbf{k}_m$, these two Fermi seas can differ in their fine-grained topological structures even through $\chi_F$ remains unchanged. The fine-grained topological structures clearly extend beyond the description provided by $\chi_F$.

\begin{figure}[!t]
\centering
\includegraphics[width=1.0\columnwidth]{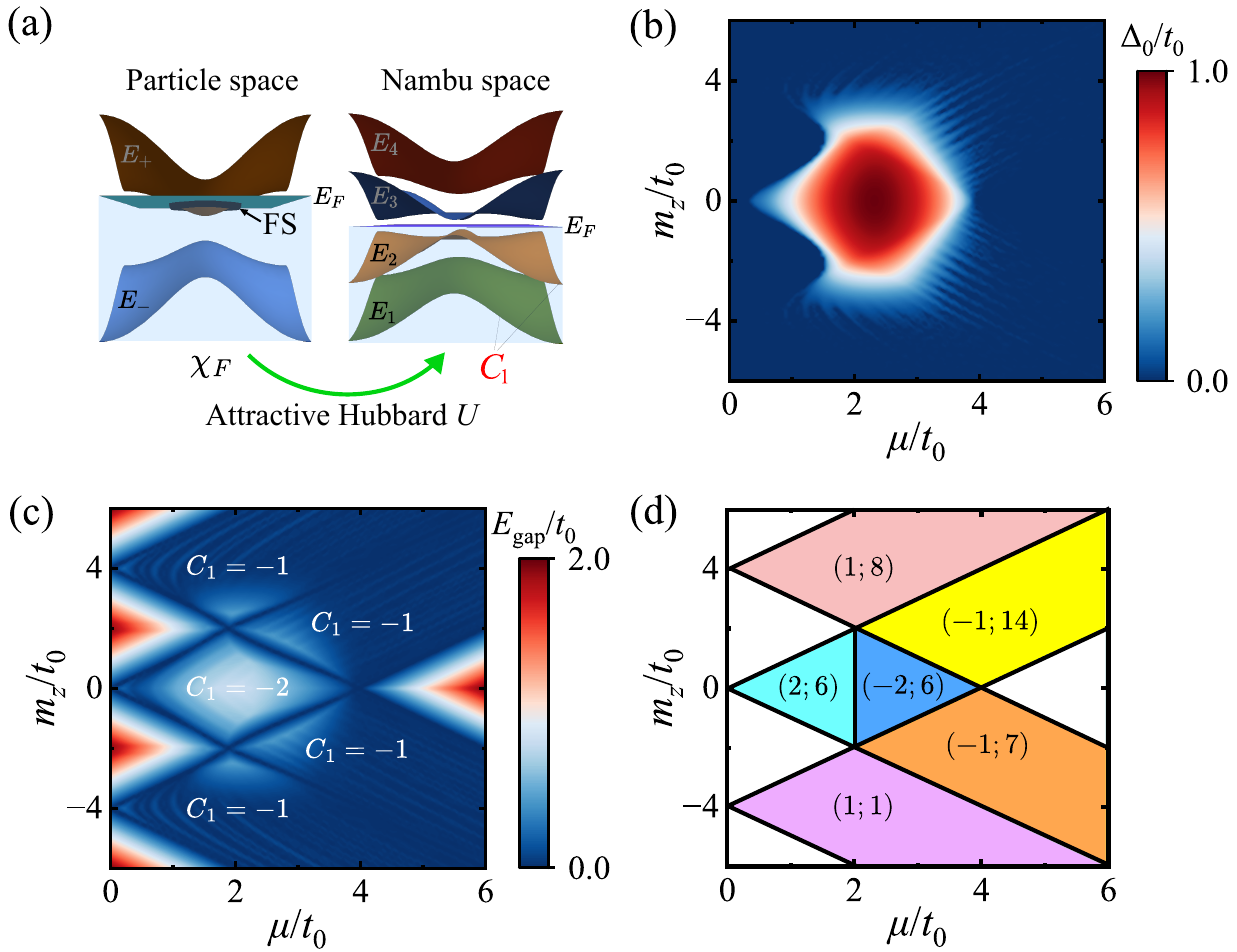}
\caption{(a) Chiral topological SC phases are induced by considering an attractive Hubbard interaction of electrons on the FS. (b)-(c) Self-consistent calculations of the pairing order for (b) and the SC bulk gap for (c). (d) Phase diagram indicated by $(\chi_F;g)$. In white regions, the system is a insulator phase. The other parameters are $t_{\text{so}}=t_0$ and $U=5t_0$.}
\label{fig:2}
\end{figure}

\begin{figure*}[!t]
\centering
\includegraphics[width=1.95\columnwidth]{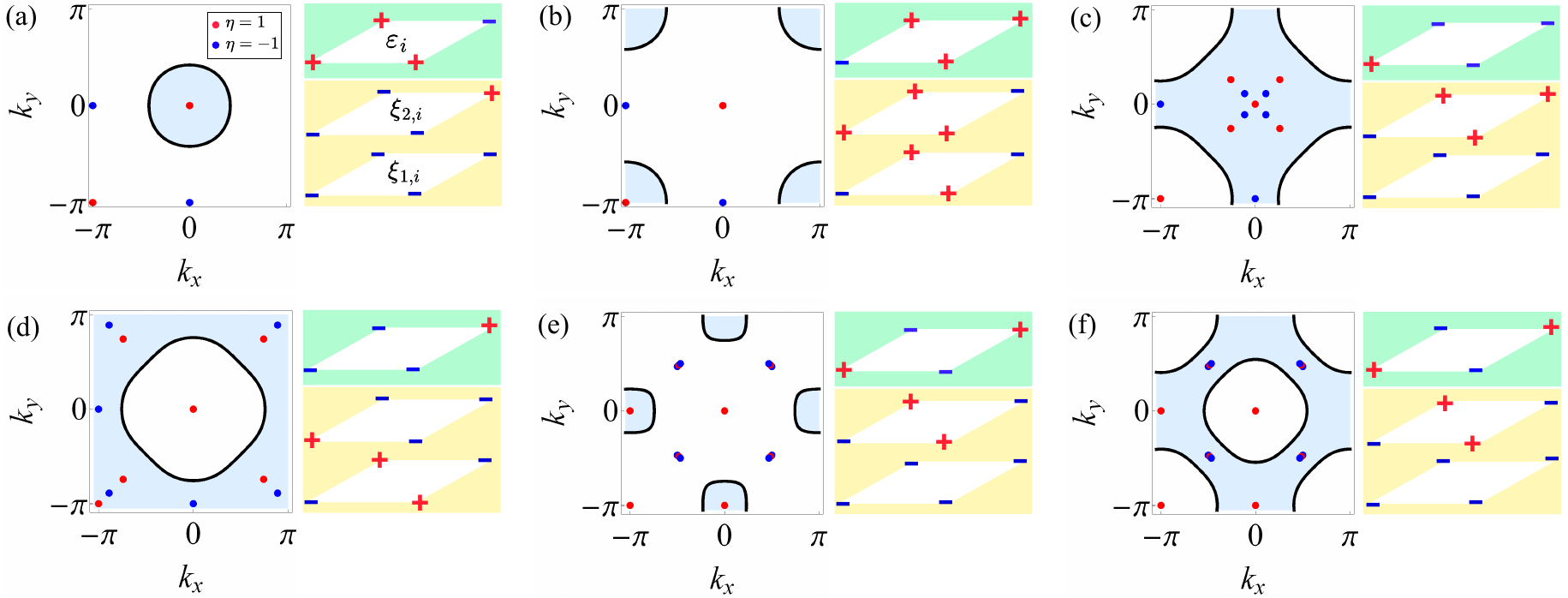}
\caption{The different configurations of Fermi seas (left) and the signs of $\varepsilon_i$ and $\xi_{j,i}$ (right). The Fermi sea topologies are characterized by $(\chi_F;g)=(1;8)$, $(1;1)$, $(-1;14)$, $(-1;7)$, $(2;6)$, and $(-2;6)$ for (a)-(f), respectively. The corresponding chiral topological SC phases are characterized by $(C_1;g_1,g_2)=(-1;15,7)$, $(-1;9,8)$, $(-1;15,1)$, $(-1;9,14)$, $(-2;15,9)$, and $(-2;15,9)$ for (a)-(f), respectively. The parameters of $(m_z,\mu,\Delta_0)$ are taken as $(3t_0,2t_0,0.249t_0)$ for (a), $(- 3t_0,2t_0,0.249t_0)$ for (b), $(2.1t_0,3t_0,0.4103t_0)$ for (c), $(-2.1t_0,3
t_0,0.4103t_0)$ for (d), $(0.2t_0,1.5t_0,0.7912t_0)$ for (e), and $(0.2t_0,2.5t_0,0.9834t_0)$ for (f), respectively. The other parameters are $t_{\text{so}}=t_0$ and $U=5t_0$.}
\label{fig:3}
\end{figure*}

To exactly capture such fine-grained topological structures, we redefine $\chi_F$ based on Morse theory. First, we assume that, in addition to having $M$ critical points in the Fermi sea, there are also $N$ critical points that are not within it. All the critical points are denoted as ${\mathbf{k}_1, \mathbf{k}_2, \cdots, \mathbf{k}_Q,\cdots,\mathbf{k}_{M+N}}$. When the parameters of the system are adjusted, the positions of certain critical points in the BZ remain unchanged, while others may shift. These can be called fixed critical points (FCPs) and adjustable critical points (ACPs), respectively. We adopt a unified index such that $i=1,\dots,Q$ are for the FCPs and $i=Q+1,\dots,M+N$ are for the ACPs. Note that the existence of FCPs requires the system to have certain symmetry, such as time-reversal symmetry or crystalline symmetry. Second, we introduce another signature 
\begin{equation}
\varepsilon_i=\text{sgn}(E_{\mathbf{k}_{i}}-E_F) 
\end{equation}
for each $\mathbf{k}_i$, where $\varepsilon_i=1$ ($-1$) describes $\mathbf{k}_i$ being located outside (inside) the Fermi sea. By defining $\kappa_i=\eta_i\varepsilon_i$ to characterize the overall signature of $\mathbf{k}_i$, the topological properties of FCPs and ACPs can be determined by the $\mathbb{Z}$-classified topological indexes $w$ and $v$, respectively. The explicit expressions are given by 
\begin{equation}
\begin{split}
w=-\frac{1}{2}\sum^Q_{i=1}\kappa_i,~~v=-\frac{1}{2}\sum^{M+N}_{i=Q+1}\kappa_i.
\end{split}
\end{equation}
Furthermore, the LTs that keep $\chi_F$ unchanged have two classes: those driven solely by FCPs crossing $E_F$, and those driven by FCPs and ACPs crossing $E_F$ simultaneously. With this, one can define a structural resolution factor
\begin{equation}
\begin{split}
g=\sum^Q_{i=1}2^{i-1}\gamma_i
\end{split}
\end{equation}
to determine the sequence properties of $\{\varepsilon_i\}$ for all the FCPs, where $(-1)^{\gamma_i}=\varepsilon_i$ and $\gamma_i\in \{0,1\}$. The ordering of FCPs is fixed once and for all according to a predefined convention, and thus $g$ is an encoding of the sequence $\{\varepsilon_i\}$ under this fixed ordering. A different ordering corresponds merely to relabeling the binary digits and does not affect the underlying physics. Finally, the Euler characteristic with intrinsic structure is defined as 
\begin{equation}
\chi_F=w+v,
\end{equation}
where $g\in \mathbb{N}$ characterizes the fine-grained topological structures of the Fermi sea. Accordingly, the topological indices $(\chi_F;g)$ provide a comprehensive description of the $2$D Fermi sea topology, as shown in Fig.~\ref{fig:1}(a). 

Besides, a schematic in Fig.~\ref{fig:1}(b) is shown to illustrate how $(\chi_F;g)$ work, in which we assume four cases of 2D Fermi sea: I, II, III, and IV. By changing the parameters of the system, case I can be sequentially deformed into the cases II, III, and IV. The fine-grained topological structures of the Fermi sea are markedly different for the cases I (II), III, and IV, even though the same $\chi_F$ are shared. More numerical examples have been provided in the Supplemental Material~\cite{SuppInfo}, showing that our theory is an effective and universal framework to fully characterize the Fermi sea topology. 

\section{Interaction-induced superconducting phases}~For a $2$D metallic system, the electrons on FSs can produce Cooper pairs under an attractive Hubbard interaction, resulting in the emergence of SC phases~\cite{PhysRevB.82.184516,bernevig2013topological,sato2017topological,chan2017generic}. Accordingly, the SC properties 
can be greatly affected by the fine-grained Fermi sea topology of their parent filled bands. Here, we consider a family of $2$D metallic bands 
\begin{equation}
E_{\pm}=\pm\sqrt{h_{1,\mathbf{k}}^2+h_{2,\mathbf{k}}^2+h_{3,\mathbf{k}}^2}-\mu,
\label{Emet} 
\end{equation}
which are from the Hamiltonian $H_0=\sum_\mathbf{k}\psi^{\dagger}_\mathbf{k}H_\mathbf{k}\psi_\mathbf{k}$ with $H_\mathbf{k}=h_{1,\mathbf{k}} \sigma_x+h_{2,\mathbf{k}}\sigma_y+h_{3,\mathbf{k}}\sigma_z-\mu$. Here $\psi_\mathbf{k}=(c_{\mathbf{k},\uparrow},c_{\mathbf{k},\downarrow})$ is the spinor in particle space and $\mu$ is the chemical potential. By taking a reasonable parameter for $\mu$, the metallic phases emerge, where the upper band $E_+$ is partially filled. Next, we introduce the attractive Hubbard interaction with the strength $U$ into this metallic system. The SC phases are induced by $U >0$, as shown in Fig.~\ref{fig:2}(a). The total Hamiltonian is written as 
$\mathcal{H}_\mathbf{k}=H_0-U\sum_\mathbf{i}n_{\mathbf{i}\uparrow}n_{\mathbf{i}\downarrow}$,
where $n_{\mathbf{i}\sigma}=c^{\dagger}_{\mathbf{i},\sigma}c_{\mathbf{i},\sigma}$ is the particle number operator. The possible pairing order is considered as $\Delta_0=\frac{U}{2P}\sum_\mathbf{k}\langle c_{\mathbf{k}\uparrow}c_{-\mathbf{k}\downarrow}\rangle$, with zero center-of-mass momentum of Cooper pairs and $P$ being the number of sites. Under the mean field approach, the interaction term is decoupled into $H_{U}=\sum_\mathbf{k}\Delta_0c^{\dagger}_{\mathbf{k}\uparrow}c^{\dagger}_{-\mathbf{k}\downarrow}+\text{h.c.}$ and $\mathcal{H}_\mathbf{k}$ can be rewritten as
\begin{equation}
\mathcal{H}_\text{MF}=\frac{1}{2}\sum_\mathbf{k}\Psi^\dagger_\mathbf{k}H_\text{MF}\Psi_\mathbf{k},~H_\text{MF}=
\begin{pmatrix}
H_\mathbf{k}&\Delta\\
\Delta^\dagger&-H^\text{T}_{-\mathbf{k}}
\end{pmatrix},
\label{HMF}
\end{equation}
where $\Delta=i\Delta_0\sigma_y$ and the basis is denoted as $\Psi_\mathbf{k}=(c_{\mathbf{k},\uparrow},c_{\mathbf{k},\downarrow},c^\dagger_{-\mathbf{k},\uparrow},c^\dagger_{-\mathbf{k},\downarrow})$ in Nambu space. Here, we take $h_{1,\mathbf{k}}=2t_\text{SO}\sin k_x\cos k_y$, $h_{2,\mathbf{k}}=2t_\text{SO}\sin k_y$, and $h_{3,\mathbf{k}}=\left[m_z-2t_0(\cos k_x+\cos k_y)\right]$, where $m_z$ is the Zeeman constant, $t_0$ ($t_\text{SO}$) is the coefficient of spin-conserved (spin-flipped) hopping, and $\sigma_{x,y,z}$ are the Pauli matrices on spin space. Since the Hamiltonian $H_\mathbf{k}$ drives quantum anomalous Hall states at $\mu=0$, the mean-field Hamiltonian $\mathcal{H}_\text{MF}$ can render the 2D chiral topological SC phases with $\mu$ being increased~\cite{PhysRevLett.100.096407,PhysRevB.82.184516,liu2014realization}. We therefore expect that the properties of chiral topological SC phases can be effected by the fine-grained Fermi sea topology of their parent filled bands.

After choosing an interaction as $U=5t_0$ and performing the self-consistent calculations, we numerically obtain the pairing order in Fig.~\ref{fig:2}(b) and the SC bulk gap in Fig.~\ref{fig:2}(c) at zero temperature, of which the Chern number $C_1$ is calculated by the TKNN number~\cite{PhysRevLett.71.3697,PhysRevB.79.094504,PhysRevB.74.045125} and is confirmed by the spectrum of Wannier charge center~\cite{PhysRevB.83.035108,PhysRevB.89.115102,benalcazar2017electric}. The details are provided in the Supplemental Material~\cite{SuppInfo}. We observe that the SC bulk gap closes at four high-symmetric momentum points $\boldsymbol{\Lambda}_i=\{\mathbf{M},\mathbf{X}_1,\mathbf{X}_2,\boldsymbol{\Gamma}\}$ for $\Delta_0^2+\mu^2=(m_z-4t_0)^2$, $m_z^2$, and $(m_z+4t_0)^2$, respectively. It is surprising that there are bulk gap closings to distinguish two topological SC phases with the same Chern number, as shown in Fig.~\ref{fig:2}(c). This result implies that these two SC phases have different fine-grained topological structures, and thus they cannot be connected through the deformations that do not close the bulk gap. By calculating $(\chi_F;g)$ of parent filled bands and showing the result in Fig.~\ref{fig:2}(d), we demonstrate that the above results are closely associated with the fine-grained Fermi sea topology of their parent filled bands. Namely, the difference of structural resolution factor $g$ in the Fermi sea determines the emergence of topological phase transitions of SC phases. For the example of $m_z>0$, the SC phases with $C_1=-1$ are in the regions of $(\chi_F;g)=(1;8)$ and $(-1;14)$, holding different $g$. Thus the bulk gap closing appears in these two SC phases. In contrast, there is no bulk gap closing for the SC phases with $C_1=-2$, since their parent filled bands give $(\chi_F;g)=(2;6)$ and $(-2;6)$ which share the same $g$.  

The above results reveal a fundamental physical mechanism, i.e., the interaction-induced SC phases can inherit the fine-grained Fermi sea topology of their parent filled bands. The reason is that the parity eigenvalues $\xi_{j,i}$ at four high-symmetric points $\boldsymbol{\Lambda}_i$ for two occupied SC bands $E_{j,\mathbf{k}}$ are related to the structural resolution factor $g$ of FCPs, where $j=1,2$. When taking a weak but finite Hubbard interaction, the SC pairing can be generally limited as $\Delta_0^2<|h_{3,\boldsymbol{\Lambda}_i}^2-\mu^2|$ at each $\boldsymbol{\Lambda}_i$. Since the SC system has inversion symmetry $\mathcal{I}=\tau_z\sigma_z$, where $\boldsymbol{\tau}$ are Pauli matrices acting on the Nambu space, the parities of these two occupied SC bands satisfy
\begin{equation}
\varepsilon_i=\prod_{j=1,2}\xi_{j,i}.
\label{vs}
\end{equation}
The detailed proof is provided in Supplemental Material~\cite{SuppInfo}. Here we give six configurations of Fermi seas in Fig.~\ref{fig:3}, which are taken in different phase regions of Fig.~\ref{fig:2}(d). It is clear that $\varepsilon_i$ and $\xi_{j,i}$ at $\mathbf{k}_i=\boldsymbol{\Lambda}_i$ obey Eq.~\eqref{vs}. With this, one can define two structural resolution factors $g_j=\sum^Q_{i=1}2^{i-1}\gamma_{j,i}$, with $\gamma_{j,i}\in\{0,1\}$ and $j=1,2$, to describe the fine-gained topological structures of the chiral topological SC phases, where $(-1)^{\gamma_{j,i}}=\xi_{j,i}$ and $Q=4$. Then the topological indices $(C_1;g_1,g_2)$ characterize the chiral topological SC phases, as shown in Fig.~\ref{fig:3}. These indices effectively distinguish two topological SC phases that have the same Chern number but different fine-grained topological structures~\cite{SuppInfo}. Besides, it should be noted that the Eq.~\eqref{vs} drives $\gamma_i=(\gamma_{1,i}+\gamma_{2,i})~\text{mod}~2=\gamma_{1,i}\oplus\gamma_{2,i}$, where $\oplus$ denotes an XOR function with the properties of $A \oplus~0=A$ and $A \oplus A=0$. Then we have 
\begin{equation}
g=\sum^Q_{i=1}2^{i-1}(\gamma_{1,i}\oplus\gamma_{2,i})=g_{1}\oplus_\text{bit} g_{2}.
\label{gg}
\end{equation}
Here $\oplus_\text{bit}$ defines a bitwise XOR function after rewriting $g$ and $g_{j}$ as binary forms. This implies that two interaction-induced topological SC phase have the different $g_j$, when their parent filled bands host the different $g$. Hence Eq.~\eqref{gg} directly confirms that the fine-grained Fermi sea topology of parent filled bands can be inherited by the topological SC phases.

\section{Anomalous Majorana interface states}~For chiral topological insulators or chiral topological superconductors, the gapless states can appear at the interface between topologically distinct systems and manifest as chiral modes, when changing the topological invariants~\cite{rosen2017chiral,zhao2023creation,zhang2024manipulation}. Especially for two distinct 2D chiral topological SC systems, the number of chiral Majorana modes at the interface is determined by $
N_\text{chiral}=|C_{1,\text{L}}-C_{1,\text{R}}|$~\cite{liu2016large,wan2024topological,yan2024rules,PhysRevB.111.L161410}, where $C_{1,\text{L}}$ and $C_{1,\text{\textsc{R}}}$ are the Chern numbers of left-hand and right-hand topological systems, respectively. Therefore, no Majorana mode appears at the interface when $C_{1,\text{L}}=C_{1,\text{R}}$.

\begin{figure}[!t]
\centering
\includegraphics[width=1.0\columnwidth]{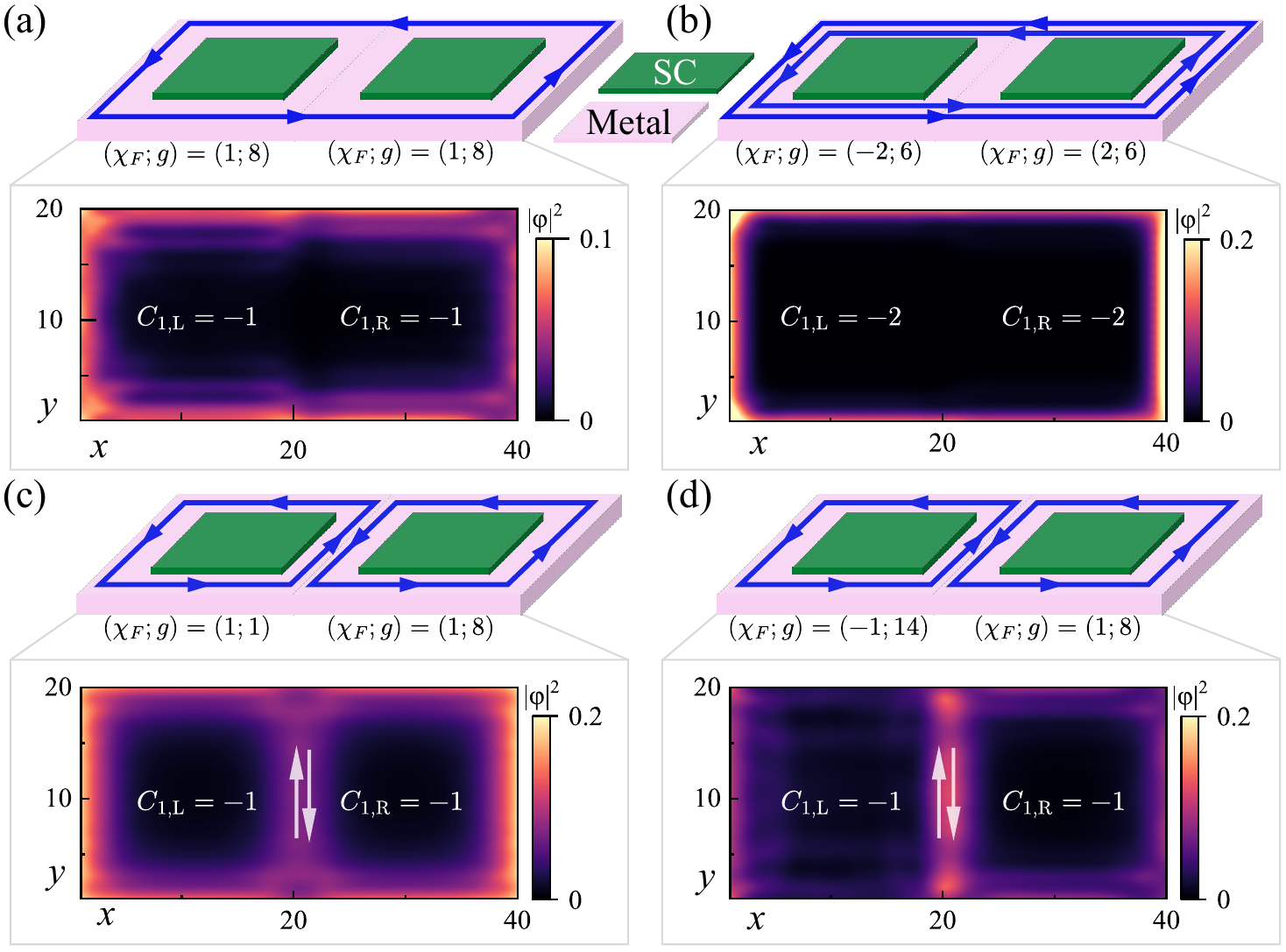}
\caption{Two connected metal/superconductor heterojunctions without (a)-(b) and with anomalous Majorana interface states (c)-(d), where the insets are the real-space distributions of 40 states near zero energy. The Chern numbers of these chiral topological SC phases are $C_{1,\text{L}}=C_{1,\text{R}}=-1$ for (a), (c), (d), and $C_{1,\text{L}}=C_{1,\text{R}}=-2$ for (b). We take $(m_{z,\text{L}},\mu_{\text{L}},m_{z,\text{R}},\mu_{\text{R}})=(3t_0,2.5t_0,3t_0,2t_0)$ for (a), $(0.2t_0,1.5t_0,0.2t_0,2.5t_0)$ for (b),$(-3t_0,2t_0,3t_0,2t_0)$ for (c), and $(2.1t_0,3t_0,3t_0,2t_0)$ for (d), respectively. The other parameters are $t_\text{so}=t_0$ and $U=5t_0$. }
\label{fig:4}
\end{figure}

However, we hereby show an anomalous phenomenon, i.e., two Majorana modes with opposite chirality can emerge in the interface of two metal/superconductor heterojunctions (MSCHs) with the same Chern numbers. This nontrivial result is associated with the difference of fine-grained Fermi sea topologies of two metals in the MSCHs. Specifically, each MSCH is produced by coupling an $s$-wave superconductor to a metal~\cite{PhysRevLett.90.137003,chung2011topological}, as shown in Fig.~\ref{fig:4}. If the band dispersion of the metal is given by Eq.~\eqref{Emet} obeying $H_\mathbf{k}$, this metal can be translated into a chiral topological superconductor due to the proximity effect of the $s$-wave superconductor and presents the chiral Majorana modes described by $\mathcal{H}_\text{MF}$~\cite{PhysRevB.82.184516}. When two metals in the connected MSCHs have different fine-grained Fermi sea topology, two topological SC phases have different structural resolution factors $g_{1,2}$ and cannot smoothly deform into each other, despite having the same Chern number. Accordingly, the Majorana interface states shall inevitably appear at the interface of two connected MSCHs.

We first take the Fermi sea topologies of two metals as $(\chi_F;g)=(1;8)$, as shown in Fig.~\ref{fig:4}(a). Both topological SC phases exhibit $(C_1;g_1,g_2)=(-1;15,7)$. Hence there is no Majorana mode at the interface of two connected MSCHs. Similar results are also observed by changing two metals into $(\chi_F;g)=(-2;6)$ and $(2;6)$, where two topological SC phases have $(C_1;g_1,g_2)=(-2;15,9)$, as shown in Fig.~\ref{fig:4}(b). In contrast, when separately choosing $(\chi_F;g)=(1;1)$ and $(1;8)$ for two metals, the different fine-grained structures of Fermi sea topology lead to $(C_1;g_1,g_2)=(-1;9,8)$ and $(-1;15,7)$ for the left-hand and right-hand MSCHs, respectively. Even though two topological SC phases still host the same $C_1$, the Majorana states emerge at the interface of two MSCHs due to the difference of $g_{1,2}$, as shown in Fig.~\ref{fig:4}(c). Similar results are also observed in Fig.~\ref{fig:4}(d), where the Fermi sea topologies of two metals are $(\chi_F;g)=(-1;14)$ and $(1;8)$. The two topological SC phases have $(C_1;g_1,g_2)=(-1;15,1)$ and $(-1;15,7)$ for the left-hand and right-hand MSCHs, respectively. These results demonstrate that the fine-grained topological structures of Fermi sea can yield interesting physical phenomena.   

\section{Discussion and Conclusion}~Our theory can apply to the 2D metallic systems possessing symmetry-protected FCPs in momentum space. Such FCPs naturally arise from time-reversal symmetry or other crystalline symmetry, such as inversion, mirror, and rotation, where $E_\mathbf{k}$ is an odd or even function of $k_i$, with $i$ taking a certain direction or both of the two directions. For generic metallic bands without any FCPs, the present structural resolution factor $g$ is no longer well defined, and extending the framework to such systems remains an interesting future problem. Besides, we emphasize that the present theory is formulated within the standard Morse-function regime, where the band dispersion remains a Morse function and the set of Fermi critical points is invariant under continuous deformations. If there is the creation or annihilation of critical-point pairs by changing the parameters of the system, it requires passing through non-Morse singularities and therefore lies outside the scope of the present framework. Within this well-defined parameter space, $g$ provides a rigorous topological index beyond the Euler characteristic.

In summary, we have demonstrated that metallic systems can host fine-grained topological structures in their Fermi seas, which go beyond the description provided by the Euler characteristic. Moreover, we have proposed a structural resolution factor to effectively characterize them, showing that such fine-grained Fermi sea topology can induce several physical phenomena: (i) the Fermi seas with matching Euler characteristics but differing structural resolution factors cannot be adiabatically connected; (ii) the interaction-induced superconducting phases can definitely inherit the fine-grained Fermi sea topology of their parent filled bands, imprinting it onto the topology of pairing states; (iii) the differences of fine-grained Fermi sea topology induce anomalous gapless states in the interface of two connected metal/superconductor heterojunctions. These nontrivial results provide an insight into the intricate topology of the Fermi sea and are expected to broad research into the Fermi sea geometry.

\section*{Acknowledgements}~We thank Long Zhang for helpful discussions. This work is supported by the National Natural Science Foundation of China (Grants No. 12404318, No. 12247101, and No. 92576202), the Fundamental Research Funds for the Central Universities (Grant No. lzujbky-2024-jdzx06), and the Natural Science Foundation of Gansu Province (Grants No. 22JR5RA389, No. 26RCKA011, and No. 25JRRA799).

\pagebreak
\clearpage
\onecolumngrid
\flushbottom
\begin{center}
\textbf{\large Supplemental Material for ``Fine-grained topological structures hidden in the Fermi sea"}
\end{center}
\setcounter{equation}{0}
\setcounter{figure}{0}
\setcounter{table}{0}
\makeatletter
\renewcommand{\theequation}{S\arabic{equation}}
\renewcommand{\thefigure}{S\arabic{figure}}
\renewcommand{\bibnumfmt}[1]{[S#1]}
\renewcommand{\citenumfont}[1]{S#1}

In this Supplemental Material, we provide the details of the fine-grained Fermi sea topology and the properties of chiral topological superconducting (SC) phases. Besides, more numerical results are also provided.

\subsection{More instructions of fine-grained Fermi sea topology}

Our theory applies to more generic $2$D metallic system with $n$ filled bands, i.e., $E_{j,\mathbf{k}}$ with $j=1,2,\cdots,n$, where only one top band $E_{n,\mathbf{k}}$ crosses Fermi level $E_F=0$. This top band is denoted as $E_\mathbf{k}$ and is used to determine the Fermi sea topology, since all the completely filled bands in the below of $E_F$ exhibit $\chi_F=0$. Furthermore, one can define a quantity 
\begin{equation}
G_\mathbf{k}=\sqrt{E_\mathbf{k}^2+\mathbf{v}_\mathbf{k}^2}
\end{equation}
and find $G=\text{min}(G_\mathbf{k})=0$ if there is a LT. And then, the certain fixed critical points (FCPs) are located at Fermi surfaces (FSs), giving a solution of $\mathbf{k}$ for $E_\mathbf{k}=\mathbf{v}_k=0$. Besides, it should be noted that the metallic bands should possess certain symmetry, such as time-reversal symmetry or crystalline symmetry, and thus the number and position of FCPs in Brillouin zone (BZ) are unchanged when the parameters of system are adjusted. Such case requires $E_\mathbf{k}$ be an odd or even function of $k_i$, where $i$ takes one or two directions of momenta.

\begin{figure*}[!b]
\centering
\includegraphics[width=1.0\columnwidth]{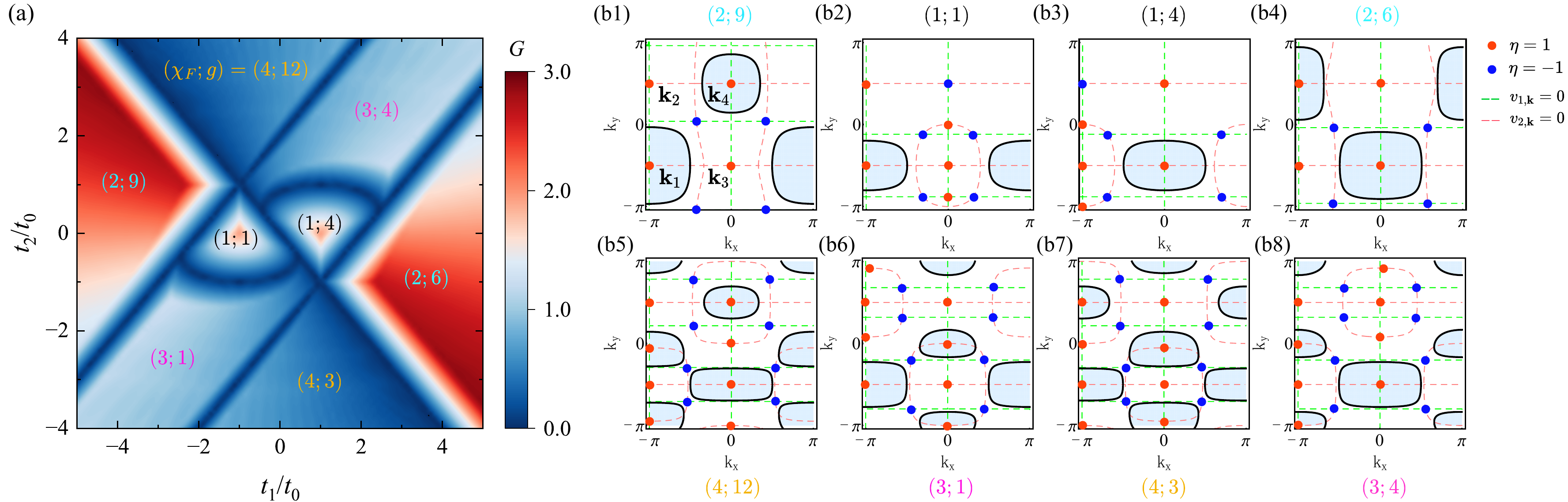}
\caption{The numerical results of $G$ for (a) and the different Fermi sea structures for (b). The black curves give the Fermi surfaces (FSs) and the lightblue regions denote the insides of Fermi seas with $E_\mathbf{k}<0$. The green and red dashed lines give $v_{1,\mathbf{k}}=0$ and $v_{2,\mathbf{k}}=0$, respectively. The critical points are captured by $v_{1,\mathbf{k}}=v_{2,\mathbf{k}}=0$, where the FCPs are $\mathbf{k}_1$, $\mathbf{k}_2$, $\mathbf{k}_3$, and $\mathbf{k}_4$, and the others are the ACPs. Here the parameters of $(t_1,t_2)$ are taken as $(-4t_0,0.5t_0)$ for $(\chi_F;g)=(2;9)$, $(-t_0,0.5t_0)$ for $(\chi_F;g)=(1;1)$, $(t_0,0.5t_0)$ for $(\chi_F;g)=(1;4)$, $(4t_0,0.5t_0)$ for $(\chi_F;g)=(2;6)$, $(-t_0,2t_0)$ for $(\chi_F;g)=(4;12)$, $(-t_0,-3t_0)$ for $(\chi_F;g)=(3;1)$, $(t_0,-2t_0)$ for $(\chi_F;g)=(4;3)$, and $(t_0,3t_0)$ for $(\chi_F;g)=(3;4)$, respectively.}
\label{fig:s1}
\end{figure*}

Unlike the example in main text where the 2D metallic band has a inversion symmetry, i.e., $E_\mathbf{k}=E_{-\mathbf{k}}$, we hereby consider a 2D typical metallic band $E_\mathbf{k}$ with mirror symmetry. This $E_\mathbf{k}$ is neither an odd nor an even function of $\mathbf{k}$, i.e., $E_\mathbf{k}\neq E_{-\mathbf{k}}\neq -E_{-\mathbf{k}}$. The corresponding band dispersion is given by
\begin{equation}
\begin{split}
E_{\mathbf{k}}=t_0+t_0\sin k_y+t_1\cos k_x\sin k_y+t_2\cos k_x\cos 2k_y.
\end{split}
\end{equation}
One can find $E_\mathbf{k}$ satisfies $E_{k_x,k_y}=E_{-k_x,k_y}$, and thus there are four FCPs $\mathbf{k}_1=(-\pi,-\pi/2)$, $\mathbf{k}_2=(-\pi,\pi/2)$, $\mathbf{k}_3=(0,-\pi/2)$, and $\mathbf{k}_4=(0,\pi/2)$ in the BZ. We numerically obtain $G$ in Fig.~\ref{fig:s1}(a) and the configurations of Fermi seas in Fig.~\ref{fig:s1}(b). It is seen that two Fermi seas with the same Euler characteristics $\chi_F$ are clearly separated by the LT. This implies that these two Fermi seas host the different fine-grained topological structures, determined by the different structural resolution factor $g$. For instance of two Fermi seas with $(\chi_F;g)=(1;1)$ and $(1;4)$, when we change the Fermi sea topology from $(\chi_F;g)=(1;1)$ into $(1;4)$, the signature of $\eta$ of $\mathbf{k}_2$ ($\mathbf{k}_4$) is changed from $\eta_2=-1$ ($\eta_4=1$) into $\eta_2=1$ ($\eta_4=-1$). The signature of $\varepsilon$ of $\mathbf{k}_1$ ($\mathbf{k}_3$) is changed from $\varepsilon_2=-1$ ($\varepsilon_3=1$) into $\varepsilon_2=1$ ($\varepsilon_3=-1$). Correspondingly, $g$ is changed from $g=1$ into $g=4$, giving the different fine-grained Fermi sea topology. These results further show that $(\chi_F;g)$ can effectively characterize the comprehensive topology of Fermi sea.

\subsection{Detailed derivations of Eq.~(8) and Eq.~(9) in main text}

We provide a detailed derivation of Eq.~(8) in main text, i.e.,
$\varepsilon_i=\prod_{j=1,2}\xi_{j,i}$. We once again write the mean-field Hamiltonian as
\begin{equation}
H_{\rm MF}(\mathbf{k})
=
\begin{pmatrix}
H_{\mathbf{k}} & i\Delta_0\sigma_y\\
-i\Delta_0\sigma_y & -H^{T}_{-\mathbf{k}}
\end{pmatrix}
\end{equation}
in the Nambu basis $\Psi_{\mathbf{k}}
=(
c_{\mathbf{k},\uparrow},
c_{\mathbf{k},\downarrow},
c^\dagger_{-\mathbf{k},\uparrow},
c^\dagger_{-\mathbf{k},\downarrow})^T$. The normal-state Hamiltonian reads $
H_{\mathbf{k}}=h_{1,\mathbf{k}}\sigma_x+h_{2,\mathbf{k}}\sigma_y+h_{3,\mathbf{k}}\sigma_z
-\mu$ with $h_{1,\mathbf{k}}=2t_{\rm SO}\sin k_x\cos k_y$, $
h_{2,\mathbf{k}}=2t_{\rm SO}\sin k_y$,
and $h_{3,\mathbf{k}}=m_z-2t_0(\cos k_x+\cos k_y)$. At each $\boldsymbol{\Lambda}_i=\{\mathbf{M}, \mathbf{X}_{1}, \mathbf{X}_{2},\boldsymbol{\Gamma}\}$, one can find
$h_{1,\boldsymbol{\Lambda}_i}=h_{2,\boldsymbol{\Lambda}_i}=0$ and the normal-state Hamiltonian reduces to $H_{\boldsymbol{\Lambda}_i}=m_i\sigma_z-\mu$ with $m_i=h_{3,\boldsymbol{\Lambda}_i}$. Accordingly, the mean-field Hamiltonian becomes
\begin{equation}
H_{\rm MF}(\boldsymbol{\Lambda}_i)=
\begin{pmatrix}
m_i\sigma_z-\mu & i\Delta_0\sigma_y\\
-i\Delta_0\sigma_y & -m_i\sigma_z+\mu
\end{pmatrix}.
\end{equation}
Furthermore, the inversion operator in Nambu space is $\mathcal I=\tau_z\sigma_z$, where $\tau_\alpha$ and $\sigma_\alpha$ are Pauli matrices acting in Nambu and spin spaces, respectively. At each $\boldsymbol{\Lambda}_i$, we have
\begin{equation}
\left[
H_{\rm MF}(\boldsymbol{\Lambda}_i),\mathcal I
\right]=0.
\end{equation}
This means that the eigenstates of mean-field Hamiltonain at $\boldsymbol{\Lambda}_i$ can be chosen as the simultaneous eigenstates of $\mathcal I$, i.e.,
\begin{equation}
\mathcal I
\ket{u_{j,i}}
=
\xi_{j,i}
\ket{u_{j,i}},
\end{equation}
with $\xi_{j,i}=\pm1$. And then, the mean-field Hamiltonian can be decomposed into two independent parity sectors. The positive-parity and negative-parity sector are spanned by $\psi_+=(
c_{\uparrow},
c^\dagger_{\downarrow})$ and $\psi_-=(
c_{\downarrow},
c^\dagger_{\uparrow})$ for the Hamiltonian
\begin{equation}
H_{+}(\boldsymbol{\Lambda}_i)
=
\begin{pmatrix}
m_i-\mu & \Delta_0\\
\Delta_0 & m_i+\mu
\end{pmatrix},
\qquad
H_{-}(\boldsymbol{\Lambda}_i)
=
\begin{pmatrix}
-m_i-\mu & -\Delta_0\\
-\Delta_0 & -m_i+\mu
\end{pmatrix},
\end{equation}
with the eigenvalues $\mathcal{E}_{+,\boldsymbol{\Lambda}_i}=m_i\pm \sqrt{\mu^2+\Delta_0^2}$ and $\mathcal{E}_{-,\boldsymbol{\Lambda}_i}=-m_i\pm \sqrt{\mu^2+\Delta_0^2}$, respectively.

We next evaluate the parity product of the two negative-energy BdG bands. If $|m_i|>\sqrt{\mu^2+\Delta_0^2}$, the two occupied states belong to the same parity sector. For $m_i>\sqrt{\mu^2+\Delta_0^2}$, both occupied states belong to the negative-parity sector, whereas for $m_i<-\sqrt{\mu^2+\Delta_0^2}$, both belong to the positive-parity sector. In either case, we have $\prod_{j=1,2}\xi_{j,i}=+1$. If
$|m_i|<\sqrt{\mu^2+\Delta_0^2}$, one occupied state belongs to the positive-parity sector and the other to the negative-parity sector, yielding $\prod_{j=1,2}\xi_{j,i}=-1$. Combining these two cases gives
\begin{equation}
\prod_{j=1,2}\xi_{j,i}
=
\operatorname{sgn}
\left(
|m_i|-\sqrt{\mu^2+\Delta_0^2}
\right)=\operatorname{sgn}\left(m_i^2-\mu^2-\Delta_0^2\right).
\label{eq:parity_product_SM}
\end{equation}
It is noted that the upper normal-state band at $\boldsymbol{\Lambda}_i$ is $E_{\boldsymbol{\Lambda}_i}=|m_i|-\mu$. For $\mu>0$, the FCPs $\mathbf{k}_i=\boldsymbol{\Lambda}_i$ satisfy
\begin{equation}
\varepsilon_i
=
\operatorname{sgn}
\left(
E_{\boldsymbol{\Lambda}_i}
\right)
=
\operatorname{sgn}
\left(
|m_i|-\mu
\right)=\operatorname{sgn}
\left(
m_i^2-\mu^2
\right).
\label{eq:epsilon_normal_SM}
\end{equation}
When the system has a weak but finite $U$, the pairing is generally limited by $\Delta_0^2<|m_i^2-\mu^2|$ at each $\boldsymbol{\Lambda}_i$. It is clear that Eq.~\eqref{eq:parity_product_SM} continuously reduces to
\begin{equation}
\varepsilon_i
=
\prod_{j=1,2}\xi_{j,i}.
\label{MFFS}
\end{equation}
This implies that the pairing can not induce additional band inversions at any high-symmetry points $\boldsymbol{\Lambda}_i$ during the transition from a normal metal to a SC state. Therefore, the parity configuration of the occupied SC bands naturally inherits the inside/outside configuration of the FCPs in the parent metal.

Finally, we prove the Eq.(9) in main text, i.e., $g=g_{1}\oplus_\text{bit} g_{2}$. Based on \eqref{MFFS}, we rewrite $\varepsilon_i=(-1)^{\gamma_i}$, $
\xi_{j,i}=(-1)^{\gamma_{j,i}}$, and $\gamma_i,\gamma_{j,i}\in\{0,1\}$, and obtain 
\begin{equation}
\gamma_i=(\gamma_{1,i}+\gamma_{2,i})~\text{mod}~2=\gamma_{1,i}\oplus\gamma_{2,i},
\end{equation}
where $\oplus$ denotes the exclusive-OR (XOR) operation. We substitute the above results into $g=\sum^Q_{i=1}2^{i-1}\gamma_i$, $g_{1}=\sum^Q_{i=1}2^{i-1}\gamma_{1,i}$, and $g_{2}=\sum^Q_{i=1}2^{i-1}\gamma_{2,i}$ to obtain
\begin{equation}
g=\sum^Q_{i=1}2^{i-1}\gamma_i=\sum^Q_{i=1}2^{i-1}(\gamma_{1,i}\oplus\gamma_{2,i})=g_1\oplus_\text{bit}g_2\\
\end{equation}
Consequently, the structural resolution factors satisfy
\begin{equation}
g=g_1\oplus_\text{bit} g_2,
\end{equation}
where the bitwise XOR operation $\oplus_\text{bit}$ acts independently on the binary digits associated with the ordered set of FCPs. For example of the chiral topological SC phase with $(C_1;g_1,g_2)=(-1;15,7)$ inheriting the fine-grained Fermi sea topology of its parent metallic bands with $(\chi_F;g)=(1;8)$, we have 
$8=(1000)_2=(1111)_2\oplus_\text{bit}(0111)_2=15\oplus_\text{bit} 7$. The similar result
can be found in the other chiral topological SC phases.

\begin{figure*}[!b]
\centering
\includegraphics[width=1.0\columnwidth]{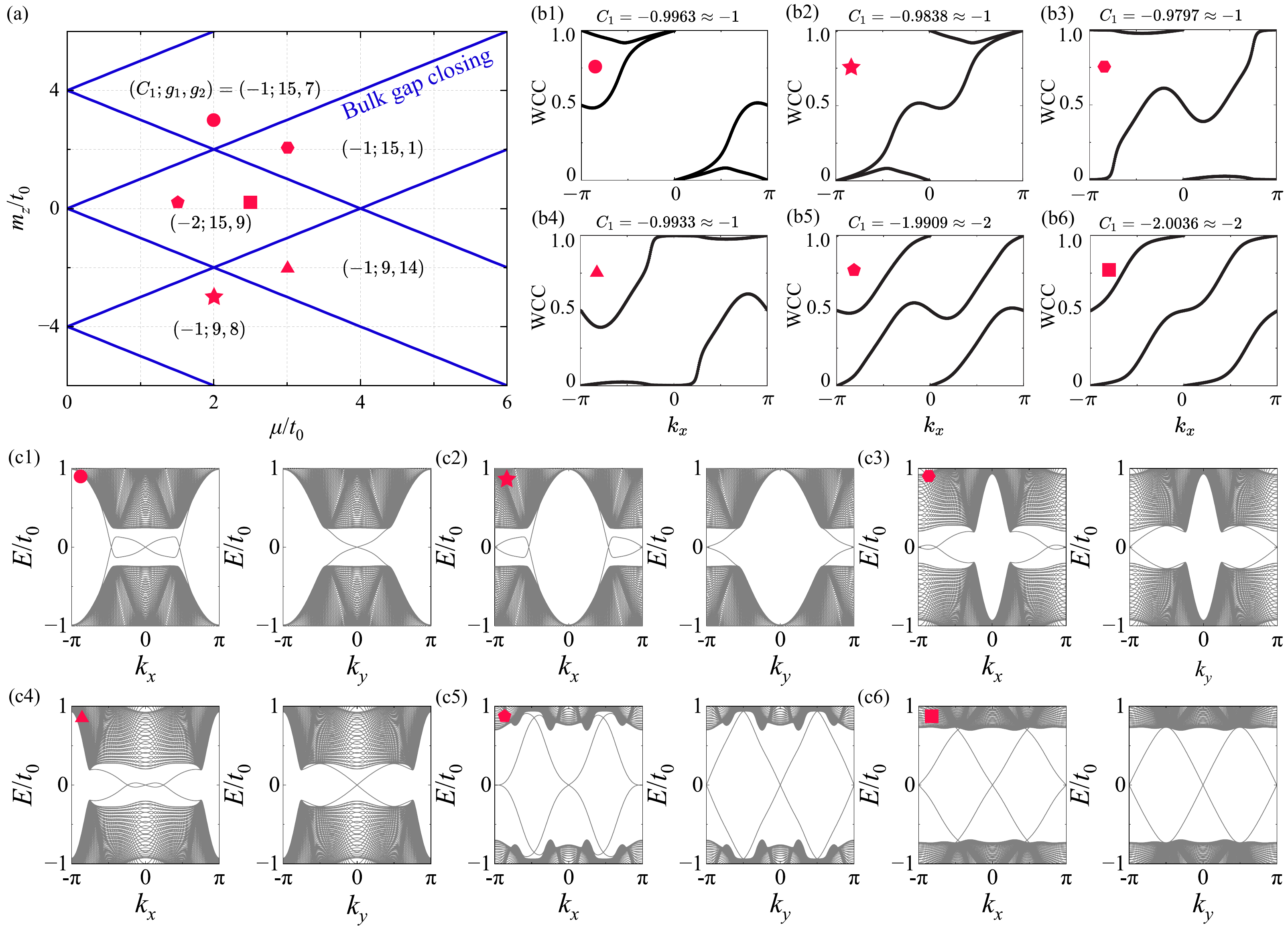}
\caption{(a) The phase diagram determined by $(C_1;g_1,g_2)$. (b1)-(b6) The numerical results of $C_1$ and the spectra of WCC, where the parameters are taken as $(m_z,\mu,\Delta_0)=(3t_0,2t_0,0.249t_0)$ for (b1), $(-3t_0,2t_0,0.249t_0)$ for (b2), $(2.1t_0,3t_0,0.4103t_0)$ for (b3), $(-2.1t_0,3t_0,0.4103t_0)$ for (b4), $(0.2t_0,1.5t_0,0.7912t_0)$ for (b5), and $(0.2t_0,2.5t_0,0.9834t_0)$ for (b6), respectively. (c1)-(c6) The energy spectra of $k_x$ or $k_y$ open boundary conditions correspond to the (b1)-(b6). The other parameters are $t_\text{SO}=t_0$ and $U=5t_0$.}
\label{fig:s2}
\end{figure*}

\subsection{More properties of chiral topological superconducting phases}

For the 2D chiral topological SC phases induced by attractive Hubbard interaction in metallic bands, it is noted that the bulk gap of $H_\text{MF}$ is closing at four high-symmetric momentum points $\mathbf{M}$, $\mathbf{X}_{1,2}$, and $\boldsymbol{\Gamma}$, when the parameters satisfy $\Delta_0^2+\mu^2=(m_z-4t_0)^2$, $m_z^2$, and $(m_z+4t_0)^2$, respectively. The corresponding results are shown in Fig.~\ref{fig:s2}(a). Generally, the topological phase transition caused by the band gap closing is accompanied by a change in the topological number. However, here we emphasize that the difference of fine-grained topology can induce the band gap closing even though two topological SC phases share the same topological number. The main reason is that the topological SC phases inherit the fine-grained Fermi sea topology of their normally filled bands. And then, 
the SC phases are characterized by $(C_1;g_1,g_2)$, as shown in Fig.~\ref{fig:s2}(a). The Chern number $C_1$ of two occupied bands is calculated by
\begin{equation}
C_1=-\frac{1}{2\pi}\iint_{\text{BZ}}dk_xdk_y\text{Im}\sum^2_{a=1}\sum^{4}_{b=3}\frac{\left<u_{a}|\partial_{k_x}H_\text{MF}|u_{b}\right>\left<u_{b}|\partial_{k_y}H_\text{MF}|u_{a}\right>-\left<u_{a}|\partial_{k_y}H_\text{MF}|u_{b}\right>\left<u_{b}|\partial_{k_x}H_\text{MF}|u_{a}\right>}{(E_{a}-E_{b})^2},
\end{equation} 
where $H_\text{MF}\left|u_{j}\right>=E_{j}\left|u_{j}\right>$ with $j=1,2,3,4$. Moreover, the Chern number $C_1$ is further confirmed by the spectra of Wannier charge center (WCC), as shown in Fig.~\ref{fig:s2}(b). The topological phases with $C_1=-1$ $(-2)$ clearly drive a winding number $W=-1$ $(-2)$ of WCC spectrum. Finally, the spectra of $H_\text{MF}$ under $k_x$ or $k_y$ open boundary conditions are provided in Fig.~\ref{fig:s2}(c), showing the difference of edge states for the same chiral topological SC phases. Since we generalize the characterization framework of the fine-grained Fermi sea topology into the fine-grained topological structures of SC phases, this demonstrates the universality of this theory. 

\end{document}